\documentclass[10pt]{article}
\usepackage{cite}
\usepackage{epsfig}

\begin{document}

\title{Reconstructions of diamond (100) and (111) 
surfaces: Accuracy of the Brenner 
potential
}

\author{A.V. Petukhov\thanks{On leave from the Institute of Crystallography, 
117234 Moscow, Russia; e-mail: {\tt petukhov@sci.kun.nl}, URL: 
{\tt www.sci.kun.nl/tvs/people/petukhov}} (a,b), A. Fasolino (a)\\
\it (a) Research Institute for Materials, 
Institute of Theoretical Physics, \\
\it University of Nijmegen, 6525 ED Nijmegen, The Netherlands \\
\it (b) Nijmegen SON Research Institute, University of Nijmegen, \\
\it 6525 ED Nijmegen, The Netherlands}

\maketitle

\begin{abstract}
We present 
a detailed comparison of the structural predictions of the 
effective many-body Brenner potential with those of \emph{ab-initio} 
studies for known reconstructions of diamond (100) and (111) 
surfaces. 
These results 
suggest high reliability of the Brenner potential for dealing 
with carbon-based structures where different types of bonding 
are present at the same time. 
\end{abstract}
\newpage

\section{Introduction}

Carbon-based structures are of a great current interest 
\cite{onions,exotic,diamondN,graphitization}.
The challenge of these systems from the fundamental point of 
view is related to the interplay between different types of 
atom bonding, leading to a uniquely large variety of 
structural phases formed by a single element, such as diamond 
and graphite, single- and multi-shell fullerenes and 
nanotubes \cite{onions} and other 
structures \cite{exotic} with many peculiar properties. 
To study the elastic properties and fracture of these structures 
and their mixtures, the transformation paths 
between them \cite{diamondN,graphitization}, etc.\ it is important 
to develop predictive schemes 
based on simplified empirical potentials, which allow large-scale 
simulations of complex structures with mixed atomic bonding, which are
often beyond the possibilities of \emph{ab-initio} calculations. 
Effective many-body empirical potentials have proven to be useful 
and predictive for a number of materials \cite{Tersoff,FurioAu,Brenner}. 
The potentials developed by Tersoff \cite{Tersoff} for group IV elements 
are very accurate for Si and Ge, also as far as interface properties are 
concerned, but less reliable for C. 

Carbon is particularly difficult for an empirical scheme 
due to the large variety of different types of C--C bonding with 
very different energetics and bond lengths $d_{\rm CC}$. For 
example, for a single C--C bond in diamond $d_{\rm CC}=1.54$ \AA,
for a conjugated bond in graphite $d_{\rm CC}=1.42$ \AA\ 
and for a double bond in H$_2$C$=$CH$_2$ $d_{\rm CC}=1.34$ \AA. 
The Tersoff potential, which has been fit to the bulk properties 
of both diamond and graphite, does not, however, distinguish 
the chemical character of the bond. At diamond surfaces, 
different types of bonding are present at the same time, leading 
to poor results of the Tersoff potential for the surface 
reconstructions as we show in detail in this work. 

Brenner \cite{Brenner} has re-parametrised the Tersoff 
potential and added nonlocal 
terms to properly account for the bond modifications induced by a 
change of bonding of neighbouring atoms. As in the Tersoff scheme, 
the potential energy of the system is written as a sum of 
effective pair terms for each bond, the energetics of which 
depends on the local environment (bond order of Tersoff) and, 
in addition, on the chemical character of the bond (single, 
double, triple or conjugated) derived by evaluating the number 
of neighbours for the atoms forming the bond and all their 
nearest neighbours. 

Diamond surfaces are an example of a rather simple system, where 
the interplay between different types of carbon bonding 
becomes important. Numerous calculations exploiting various 
\emph{ab-initio} approaches and extensive experimental data are 
available, making the diamond surfaces an important check 
point to verify the accuracy and predictive power of empirical 
schemes. The Tersoff potential for C yields the 
unreconstructed (111) $(1\times1)$ surface as the most stable 
against the experimental evidence of the $(2\times1)$ Pandey chain 
reconstruction \cite{Pandey} analogous 
to that of Si(111). For the (001) face it 
strongly favours an asymmetric re-arrangement of carbon atoms 
beneath the raw of unbuckled dimers \cite{ourPSS}. 
In the present work we compare in detail the predictions 
of the Brenner and Tersoff potentials with results of 
\emph{ab-initio} calculations 
for known reconstructions of the diamond(100) and (111) surfaces. Our 
results reveal high quantitative accuracy of the Brenner potential. 
Since the parameters are fit to the \emph{bulk} 
properties of diamond and graphite and to properties of various 
hydrocarbon molecules, the high accuracy at the \emph{surface} suggests 
a high predictive power of the potential at short distances. 
With further modifications to include also long-range 
interactions (beyond 2 \AA, the cut-off of the potential) 
\cite{Sinnott,Che}, which are now under development \cite{tobepub}, 
the Brenner potential promises to become a powerful tool 
to investigate carbon-based structures on a large scale. 

\section{(100) surface}

\begin{figure}
\hspace*{\fill}%
\includegraphics[width=0.8\linewidth]{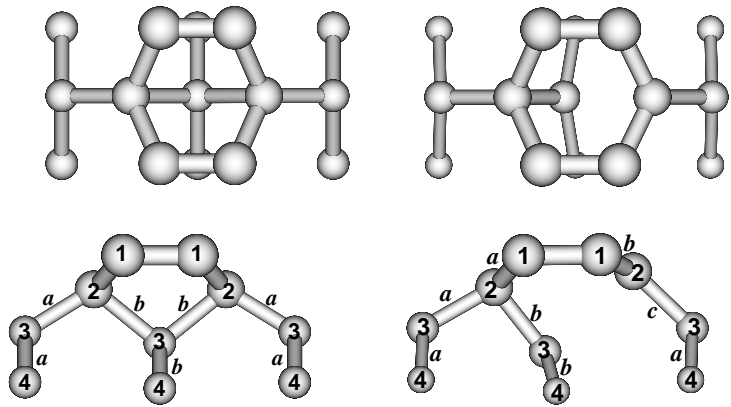}\hspace*{\fill}
\caption{Unit cell of (left) the symmetric (2$\times$1) and (right) 
the asymmetric (2$\times$1)a reconstructions of diamond(100) after 
MC energy minimisation with the Brenner potential. The top 
and bottom panels show the top (100) and the side (011) views. The 
numbers on the atoms denote the atomic layer they belong to. 
Inequivalent bonds between two atomic layers are labelled with 
letters $a$, $b$ and $c$.}
\label{100}
\end{figure}

Our previous study \cite{ourPSS} of the diamond(100) surface 
with the Tersoff potential has suggested a new 
reconstruction with a strongly asymmetric rearrangement of atoms 
in deeper layers. These predictions have now 
been verified using the off-lattice Monte Carlo (MC) technique 
with the Brenner potential. We have confirmed that 
both the symmetric $(2\times1)$ and asymmetric $(2\times1)$a 
(where `a' stands for `asymmetric') 
structures  shown in Fig.~\ref{100} correspond to local energy minima. 
However, the energies for the two reconstructions
given by the Brenner potential are found in a reversed order compared 
to the prediction of the Tersoff potential. Relative to the energy 
of the relaxed $(1\times1)$ ideal diamond (100) surface, the energy 
gain is found with the Brenner potential to be $5.40$ and 
$4.19$ eV per surface dimer for the $(2\times1)$ and $(2\times1)$a 
structures, respectively. For comparison, the Tersoff predictions 
were $0.26$ and $1.55$ eV, respectively \cite{ourPSS}. 

The length of the bonds between atoms in the top four layers are 
given for the two structures in Table~\ref{table100}. For comparison, 
we also give the results obtained with the Tersoff potential \cite{ourPSS} 
and those found in \emph{ab-initio} calculations \cite{Furthmuller96}. 
Note that, from the chemical point of view, each surface 
atom at the bulk-terminated surface has two un-paired electrons (two 
dangling bonds). Therefore, the dimer bond (bond 11 in Table~\ref{table100}) 
in the symmetric $(2\times1)$ surface structure has the 
character of a double-bond. The Brenner potential correctly 
reproduces the length of the bond 11, which is much shorter than both 
the single C--C bond in diamond and the conjugated bond in graphite, but 
rather close to the length of a double bond. 
It quantitatively agrees with the dimer bond length of 1.37 \AA\ 
for the $(2\times1)$ diamond(100) surface found in \emph{ab-initio} 
calculations \cite{Furthmuller96,Kruger95,Bechstedt94}. Conversely,
the Tersoff potential, which does not include the nonlocal terms, 
predicts very different length of the dimer bond; it also 
gives much smaller reconstruction energy for both structures 
since the chemical character of stronger double and conjugated 
bonds is not accounted for. With the Brenner potential, in the asymmetric 
$(2\times1)$a structure, the bond 11 is 
elongated up to 1.437 \AA\ (close to the graphite value) since it 
becomes a member of a 
conjugated ($\pi$-bonded) system. Note that the bond 12$b$, 
which also connects three-fold coordinated atoms, has a very 
similar length. 

\begin{table}
\caption{Bond lengths for the dimerised symmetric $(2\times1)$ and 
asymmetric $(2\times1)$a reconstructions (given in \AA). The bond 
between an atom in the $N$th layer and one in the $M$th layer is 
labelled by $NM$. Whenever more than one nonequivalent bond is present, 
they are labelled by $a$, $b$ and 
$c$ in Fig.~\ref{100} and are displayed in the table in this order. 
For completeness, we show the bond length as given by \emph{ab-initio} 
calculations \cite{Furthmuller96}, the Brenner 
[this work] and Tersoff \cite{ourPSS} empirical potentials as well 
as the corresponding bulk diamond bond lengths.}
\label{table100}
{\scriptsize
\begin{tabular}{|c||c||c|c||c|c|}
\hline 
   &\emph{ab-initio} \cite{Furthmuller96}& \multicolumn{2}{c||}{Brenner [this work]} 
  & \multicolumn{2}{c|}{Tersoff \cite{ourPSS}} \\ 
\cline{2-6}
bond&$(2\times1)$&$(2\times1)$& $(2\times1)$a       &$(2\times1)$&$(2\times1)$a 
\\ \hline
11 &1.37      &1.3807        &1.4370                &1.542       &1.487\\
12 &1.50      &1.5096        &1.4937, 1.4318        &1.515       &1.496, 1.565\\
23 &1.55, 1.57&1.5059, 1.5905&1.5339, 1.5287, 1.5321&1.524, 1.579&1.541, 1.489, 1.506\\
34 &1.56, 1.50&1.5670, 1.5140&1.5856, 1.4739        &1.570, 1.521&1.582, 1.477\\ \hline\hline
bulk&  1.53   & \multicolumn{2}{c||}{1.5407}         & \multicolumn{2}{c|}{1.5445} \\ \hline
\end{tabular}
}
\end{table}

Examination of the first two columns in Table~\ref{table100} 
reveals the high accuracy of the predictions of the Brenner 
potential for the bond lengths of the symmetric $(2\times1)$ 
diamond(100) reconstruction. 
Except a slightly higher difference of the length of the bonds 23$a$ 
and 23$b$, our results agree with the \emph{ab-initio} results 
\cite{Furthmuller96} within 0.01 \AA\ accuracy. 
Note that the \emph{ab-initio} approach of Ref.~\cite{Furthmuller96} 
underestimates the bulk bond length by 0.01 \AA\ with respect to the 
experimentally-determined value. 
In contrast to the surprisingly good quantitative agreement of 
the structural data, the reconstruction energy is different. 
The value of $5.40$ eV/dimer 
as given by the Brenner potential is found between the 
\emph{ab-initio} values (3.02 \cite{Furthmuller96}, 
3.36 \cite{Kruger95} and 3.52 \cite{Bechstedt94} eV/dimer) 
and the reconstruction energy given by the semi-empirical 
SLAB-MINDO scheme (7.86 eV/dimer \cite{Zheng91}). 

\section{(111) surface}

\begin{figure}
\hspace*{\fill}%
\includegraphics[width=0.8\linewidth]{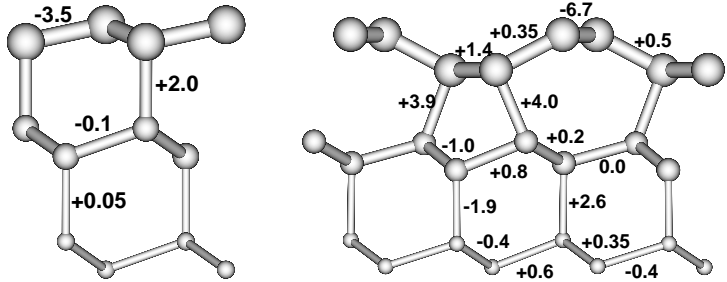}\hspace*{\fill}
\caption{Side $(\bar{1}10)$ views of (left) the relaxed $(1\times1)$ and 
(right) the Pandey $(2\times1)$ reconstruction of the (111) surface 
along with the bond 
lengths shifts relative to the bulk value as given by the Brenner potential.}
\label{111}
\end{figure}

We do not consider the Tersoff potential here since it gives the unreconstructed
$(1\times1)$ surface as the minimum energy structure. 
Fig.~\ref{111} shows the relaxed $(1\times1)$ and the $(2\times1)$ 
Pandey chain reconstructions of the diamond(111) surface as given 
by the Brenner potential. In brief, 
the most important changes of the bond lengths as compared to 
the bulk value are as follows. 
For the  relaxed $(1\times1)$ structure: 
(i) the contraction of the bond within the first bilayer by $-3.5$\%\ 
agrees with \emph{ab-initio} values of $-3.1$\%\ \cite{Vanderbilt84}, 
$-4.0$\%\ \cite{Stumpf} and $-4.2$\%\ \cite{Scholze}; (ii) 
the elongation of the bond between the first and second bilayer 
by $+2.0$\%\ agrees with $+2.1$\%\ of Ref.~\cite{Vanderbilt84} 
but seems underestimated compared to $+8.7$\%\ and $+9$\%\ 
of Refs.~\cite{Scholze,Stumpf}. In the $(2\times1)$ Pandey 
reconstruction: (iii) the $\pi$-bonded upper chain bond length 
of 1.437 \AA\ ($-6.7$\%) well compares to \emph{ab-initio} values 
of 1.47 \AA\ ($-4.4$\%) \cite{Vanderbilt84}, 1.44 \AA\ 
\cite{Tosatti} and 1.43 \AA\ ($-6.5$\%) \cite{Scholze}; 
(iv) the lower chain elongation by $+1.4$\%\ is close to 
$+0.7$\%\ \cite{Vanderbilt84} and $+0.9$\%\ \cite{Scholze}; 
(v) the stretch of the bonds between the first and second bilayers 
by $+3.9\%$ and $+4.0$\%\ seems to be underestimated with respect 
to the \emph{ab-initio} values of $+8.1$\% \cite{Vanderbilt84}, 
$+8$\%\ \cite{Tosatti}, $+4.5\%$ and $+6.6$\% \cite{Scholze}. 
Further comparison with the results of 
Refs.~\cite{Vanderbilt84,Scholze,Tosatti,Stumpf} shows that 
all other bond shifts agree with the \emph{ab-initio} 
calculations within $\sim$1\%. Therefore, except a tendency 
to underestimate the elongation of the bonds between 
the top and second bilayer, our structural results for diamond(111) 
agree remarkably well with \emph{ab-initio} 
predictions \cite{Vanderbilt84,Scholze,Tosatti,Stumpf}. 

Relative to the energy of the bulk-terminated diamond(111), 
we find energy gains per $1\times1$ unit cell 
of 0.244 eV for the $(1\times1)$ structure 
(cf 0.37 eV \cite{Vanderbilt84} and 0.57 eV \cite{Scholze})
and 1.102 eV for the Pandey reconstruction (cf 0.47 eV 
\cite{Vanderbilt84} and 1.40 eV \cite{Scholze}). 

We note that 
there is a long-standing debate on the structural and electronic 
properties of the diamond(111) surface. An important issue is whether 
this surface is metallic or semiconducting. In most 
calculations \cite{Vanderbilt84,Scholze} the band of
surface states is metallic whereas experimentally the highest occupied 
state is at least 0.5 eV below the Fermi level \cite{Graupner97}. 
Dimerisation along the $\pi$-bonded chain could open the 
surface gap but only one total-energy calculation 
obtains slightly dimerised chains yielding a 0.3 eV gap \cite{Tosatti} 
in the surface band. Experimentally, recent X-ray data \cite{X-Vlieg} 
does not show any dimerisation but favour the $(2\times1)$ 
reconstruction accompanied by a strong tilt of the $\pi$-bonded chains, 
similar to the  $(2\times1)$ reconstruction of Si(111) and Ge(111). 
The tilt is however not confirmed by theoretical 
studies \cite{Vanderbilt84,Scholze,Tosatti}. 

Neither dimerisation nor buckling of the $\pi$ chain is found in 
our results for the Pandey reconstruction in agreement to most 
\emph{ab-initio} results. However, 
our recent MC study \cite{ourPRB} of the structure of diamond(111) 
based on the Brenner potential has shown that, in addition to the stable 
$(2\times1)$ Pandey chain reconstruction, there exist additional 
meta-stable states, specific for carbon,  
with all surface atoms in three-fold graphite-like bonding. 
Since the energy of these metastable states is very close to 
the one of the Pandey $(2\times1)$, these structures can coexist with the 
Pandey structure at a real surface. Moreover, due to symmetry breaking 
induced by a strong dimerisation of the lower (4-fold coordinated) 
atomic chain in the first bilayer, the meta-stable 
reconstructions is likely to exhibit 
semiconducting behaviour. Although the new structures and their surface 
electronic properties ought to be checked in \emph{ab-initio} studies, 
the high accuracy of the Brenner potential demonstrated in this work 
strongly supports this prediction. 

\section{Conclusion}

We have performed off-lattice Monte Carlo study of the (100) 
and (111) diamond surfaces with the empirical many-body 
Brenner potential \cite{Brenner} and 
compared the results in detail with those obtained with the 
Tersoff potential \cite{ourPSS} and with \emph{ab-initio} 
approaches 
\cite{Furthmuller96,Kruger95,Bechstedt94,Vanderbilt84,Stumpf,Scholze}. 
We find that the Brenner potential is extremely accurate 
in describing the structural properties at surfaces, supporting
the recent predictions 
\cite{ourPRB} of new 
meta-stable reconstructions of diamond(111). 
On the other hand, the Tersoff potential \cite{Tersoff}, 
which does not distinguish the chemical character of the bond, 
turns out to give a poor description of surface properties. 

The Brenner potential, however, cannot describe weaker long-range 
interactions, such as the interplanar interactions in 
graphite due to the cut-off at 2 \AA. 
This is the most serious limitation 
to be overcome. Recently, further modifications 
of the Brenner potential to include also long-range 
interactions beyond 2 \AA\ are being proposed 
\cite{Sinnott,Che,tobepub}. Given the high accuracy 
of the short-range part, the modified Brenner 
potential \cite{Sinnott,Che,tobepub} promises to become a 
powerful tool to investigate carbon-based structures on a large scale. 

\paragraph{\emph{Acknowledgements}} 
We would like to thank Daniele Passerone, Furio Ercolessi and  Erio 
Tosatti for their help in implementation the off-lattice grand 
canonical Monte Carlo code. Valuable discussions with Elias Vlieg, 
Frank van Bouwelen, Rob de Groot, Hans ter Meulen, Willem van 
Enckevort and John Schermer are acknowledged.

\end{document}